\PassOptionsToPackage{cmyk,table}{xcolor}
\documentclass[conference,a4paper,flushend]{iaria} 

\pdfoutput=1 

\usepackage[babel=true,english=american]{csquotes}
\usepackage[english]{babel} 

\usepackage[shortcuts]{extdash} 

\usepackage[alwaysadjust]{paralist}
\usepackage[caption=false,font=footnotesize]{subfig}

\usepackage[
babel=true, 
expansion=alltext,
protrusion=alltext-nott, 
nopatch=eqnum, 
final 
]{microtype}
\usepackage{fontawesome} 
\usepackage{relsize}     
\usepackage{lipsum}      

\title{Stochastic Compartment Model of Epidemic Spreading in Complex Networks with Mortality and Resetting}
\author{
  \IEEEauthorblockN{%
    Thomas M. Michelitsch\ \orcidlink{0000-0001-7955-6666} and Bernard Collet }
  \IEEEauthorblockA{%
    Sorbonne Université, CNRS\\
    Institut Jean Le Rond d’Alembert\\
    F-75005 Paris, France\\
    e-mail:  {\tt$$thomas.michelitsch$$@sorbonne-universite.fr} \\ {\tt$$bernard.collet$$@sorbonne-universite.fr}     }
    \\
    \IEEEauthorblockN{%
    Michael Bestehorn \orcidlink{0000-0002-3152-8356} }
    \IEEEauthorblockA{%
    Institut f\"ur Physik\\
    Brandenburgische Technische Universit\"at Cottbus-Senftenberg\\
    Cottbus, Germany \\
    e-mail:  {\tt$$bestehorn$$@b-tu.de} 
    }
    \\
    \IEEEauthorblockN{%
    Alejandro P. Riascos \orcidlink{0000-0002-9243-3246} }  
    \IEEEauthorblockA{%
    Departamento de Física\\
    Universidad Nacional de Colombia\\
    Bogota, Colombia  \\
    e-mail:  {\tt$$alperezri$$@unal.edu.co} 
    }
    \\
    \IEEEauthorblockN{%
     Andrzej F. Nowakowski \orcidlink{0000-0002-5018-2661} }  
    \IEEEauthorblockA{%
    School of Mechanical, Aerospace and Civil Engineering\\
    University of Sheffield\\
    Sheffield, United Kingdom \\
     e-mail:  {\tt$$a.f.nowakowski$$@sheffield.ac.uk} 
    }
    }
\begin{document}
\maketitle
\begin{abstract}
We propose an epidemic compartment model, which includes mortality caused by the disease, but
excludes demographic birth and death processes. 
Individuals are represented by random walkers, which are in one of the following states (compartments) S (susceptible to infection), E (exposed: infected but not infectious corresponding to the latency period), I (infected and infectious), R (recovered, immune), D (dead). The disease is transmitted with a certain probability at contacts of I to S walkers.
The compartmental sojourn times are independent random variables drawn 
from specific (here Gamma-) distributions.
We implement this model into random walk simulations. Each walker performs an 
independent simple Markovian random walk on a graph, where we consider a Watts-Strogatz (WS) network. Only I walkers may die.
For zero mortality, we prove the existence of an endemic equilibrium for basic reproduction number ${\cal R}_0 > 1$ and for which the disease free (globally healthy) state is unstable.
We explore the effects of long-range-journeys (stochastic resetting) and mortality.
Our model allows for various interpretations, such as certain chemical reactions, the propagation of wildfires, and in population dynamics.
\end{abstract}

\section{Introduction}
Sudden outbreaks of epidemics are recurrently threatening humanity and represent major challenges for human societies and public health services.
Since the breakout of the COVID-19 pandemic, epidemic models have attracted considerable attention. More than ever, there is a need of basic understanding of the
underlying mechanisms of epidemic propagation. So-called compartmental models, where the individuals of a
population are divided according to their states of health, have become popular in the field of epidemic modeling. The first model of this type was introduced  a century ago in the seminal work of Kermack and McKendrick [1], where individuals are in one of the states (compartments) susceptible (to infection) - S, infected and infectious - I, recovered (immune) - R.
While standard SIR models are able to capture essential features of some common infectious diseases such as mumps, measles, rubella and others, they have 
revealed to be unable to describe persistent oscillatory and quasi-periodic behavior or spontaneous outbursts, features, which are observed in many cases.
One of the first works tackling the issue of oscillatory dynamics is the one by Sober [2].
The classical SIR model has been generalized in many directions 
[3]-[6] and consult [7] for a
model related to the context of COVID-19 pandemic.

\section{Related Work}
\label{related_work}
In order to relate macroscopic compartment models to microscopic dynamics, epidemic spreading  
has been studied in random graphs with emphasis on the complex interplay of the network topology and spreading features
[8]-[11]. Further works consider stochastic compartmental models combined with random walk approaches
[12]-[19] including non-exponentially distributed compartmental sojourn times leading to non-Markovian models 
[20]-[24]. An increasing number of works consider epidemic propagation on networks. In reference [19], involving generalized Laplacian operators, spreading features are thoroughly analyzed, where an upper bound for the epidemic SIS threshold for any graph topology is obtained.

In the present paper, we explore the spreading of a disease by combining a microscopic multiple random walker's approach with a compartment model exhibiting random compartmental sojourn times. Unlike most existing models, 
our point of departure are exact stochastic equations describing the transitions among the compartments 
(see (\ref{SCEIRS-D-model}) and (\ref{SCEIRS-D-model-mortality})) from which exact non-Marlovian evolution equations can be obtained by averaging over the involved random variables.
By conducting a linear stability analysis, we prove for zero mortality that the disease 
free state is stable for ${\cal R}_0 < 1$ and unstable for ${\cal R}_0 > 1$ (${\cal R}_0$ denotes the basic reproduction number), where a globally stable endemic state exists whenever the compartment sojourn times have finite means, and for which we obtain explicit formulas (see (\ref{endem})). We validate our analytical results by random walk simulations in a Watts-Strogatz (WS) network and also show some animated videos of the spreading dynamics. Related works to our model can be found in references [17], [21]-[24] and [34].

\section{Mean field compartment model}
\label{mean-field}
Here, we study the large class of infectious diseases with direct transmission among individuals, which also exhibit mortality. The large list of these diseases includes Influenza, COVID-19, Chickenpox, Hepatitis A, Ebola, and many others. We propose a compartment model, in which individuals ("random walkers") are in one of the following states (compartments) S (susceptible to infection), E (exposed: infected but not infectious corresponding to the latency period), I (infected and infectious), R (recovered, immune), and D (dead). 
We assume random waiting times $t_E,t_I,t_R$ in compartments E, I, R. The delay time $t_E$ is the latency period, i.e., the time between the moment of infection (transition S to E) and outbreak of the disease (transition E to I).
$t_I$ is the duration of the disease (infected and infectious state) during which the walker can infect S walkers and die. We introduce a random survival time $t_M$ measured from the moment of transition into compartment I (outbreak of the disease). The walker survives if $t_M>t_I$ and dies otherwise (when $t_M<t_I$).
A surviving walker passes through the SEIRS pathway
 \begin{center}
 {\Large \textcolor{blue}{S} $\to$ \textcolor{yellow}{E} $\to$ \textcolor{red}{I} $\to$ \textcolor{green}{R} $\to$ \textcolor{blue}{S}}.
 \end{center}
A walker which dies from the disease (i.e., $t_M< t_I$) runs through the SEID pathway
\begin{center}
{\Large \textcolor{blue}{S} $\to$ \textcolor{yellow}{E} $\to$ \textcolor{red}{I} $\to$ \textcolor{black}{D}}.
\end{center}
For the infection rate, we assume a simple bilinear function inspired from the mass-action law
\begin{equation}
\label{infection_law}
{\cal A}(t)= \beta S(t)J(t) ,
\end{equation}
where $\beta >0$ is a constant, which contains the information on the probability of infection in a contact of an S and I walker and
features of the random walks.
The stochastic formulation of the evolution equations of the compartmental population fractions reads
\begin{equation}
 \label{SCEIRS-D-model}
 \begin{array}{clr}
 \displaystyle \frac{d S(t)}{dt} = & - {\cal A}(t) + \big\langle {\cal A}(t-t_E-t_I-t_R)\Theta(t_M-t_I)\big\rangle & \\[1ex]
 &+J_0 \big\langle \delta(t-t_I-t_R) \Theta(t_M-t_I) \big\rangle & \\[1ex] & +  R_0\big\langle \delta(t-t_R) \big\rangle &  \\[1ex]
 \displaystyle \frac{d E(t)}{dt}  = & {\cal A}(t) - \big\langle {\cal A}(t-t_E) \big\rangle &\\[1ex] 
 \displaystyle  \frac{d J(t)}{dt}   = &\big\langle {\cal A}(t-t_E) \big\rangle  - \big\langle  {\cal A}(t-t_E-t_I) \Theta(t_M-t_I) \big\rangle  & \\[1ex]
 & \displaystyle  -  J_0  \big\langle \delta(t-t_I) \Theta(t_M-t_I) \big\rangle  - \frac{d D(t)}{dt} & \\[1ex]
 \displaystyle  \frac{d R(t)}{dt} = & \big\langle {\cal A}(t-t_E-t_I) \Theta(t_M-t_I) \big\rangle  &\\[1ex] 
 & \displaystyle +J_0  \big\langle \delta(t-t_I)\Theta(t_M-t_I) \big\rangle & \\[1ex]
 &  -J_0 \big\langle \delta(t-t_I-t_R) \Theta(t_M-t_I) \big\rangle        & \\[1ex]
 &  - \big\langle {\cal A}(t-t_E-t_I-t_R)\Theta(t_M-t_I) \big\rangle  & \\[1ex]
 &    - R_0 \big\langle \delta(t-t_R)\big\rangle    &
 \end{array}
\end{equation}
and the mortality rate
\begin{equation}
 \label{SCEIRS-D-model-mortality}
\frac{d D(t)}{dt}  = J_0 \big\langle \delta(t-t_M)\Theta(t_I-t_M)  \big\rangle  + \big\langle {\cal A}(t-t_E-t_M)\Theta(t_I-t_M) \big\rangle .
\end{equation}
$S(t), E(t), J(t), R(t), D(t)$ denote, the fractions of the susceptible, exposed, infected, recovered (immune), and 
dead walker's populations, where $S(t)+E(t)+J(t)+R(t)+D(t)=1$. We consider initial conditions 
$S(0)=S_0$, $J(0)=J_0$, $E(0)=0$, $R(0)=R_0$, $D(0)=0$ and assume that the disease occurs at $t=0$ for the first time
with a few infected walkers $J_0$, no exposed and dead walkers, and possibly some immune (vaccinated) walkers $R_0$, allowing to explore effects of vaccination.
$\Theta(..)$ indicates the Heaviside unit step function, $\delta(..)$ the Dirac's $\delta$-distribution, and $\big\langle \ldots\big\rangle$ stands for averaging with respect to the contained (independent) random variables $t_E,t_I,t_R, t_M >0$ drawn from probability density functions (PDFs) 
$$
Prob(t_{E,I,R,M}\in [\tau,\tau+{\rm d}\tau]) =  K_{E,I,R,M}(\tau){\rm d}\tau 
$$
indicating the probabilities that $t_{E,I,R,M} \in [\tau,\tau+{\rm d}\tau]$. The following averaging rule applies
\begin{equation}
\label{averag_rule}
\big\langle f(t_{E,I,R,M}) \big\rangle = \int_0^{\infty}f(\tau) K_{E,I,R,M}(\tau){\rm d}\tau .
\end{equation}
For causal functions as in (\ref{SCEIRS-D-model}) this yields convolutions
$$\big\langle A(t-t_{E,I,R,M}) \big\rangle = \int_0^tA(t-\tau)K_{E,I,R,M}(\tau){\rm d}\tau .$$
With these relations, the evolution equations
(\ref{SCEIRS-D-model}) and (\ref{SCEIRS-D-model-mortality}) can be averaged taking convolution forms (see [22, 23] for details).
\paragraph{Zero mortality -- endemic equilibrium}  
The limit of immortality of the walkers is retrieved from (\ref{SCEIRS-D-model}) for $t_M=\infty$ 
thus $\Theta(t_M-t_I)=1$ and $\Theta(t_I-t_M)=0$ and therefore $\frac{d}{dt}D(t) =0$.
Then equations (\ref{SCEIRS-D-model}) read
\begin{figure*}[t!]
\centerline{
\includegraphics[width=0.88\textwidth]{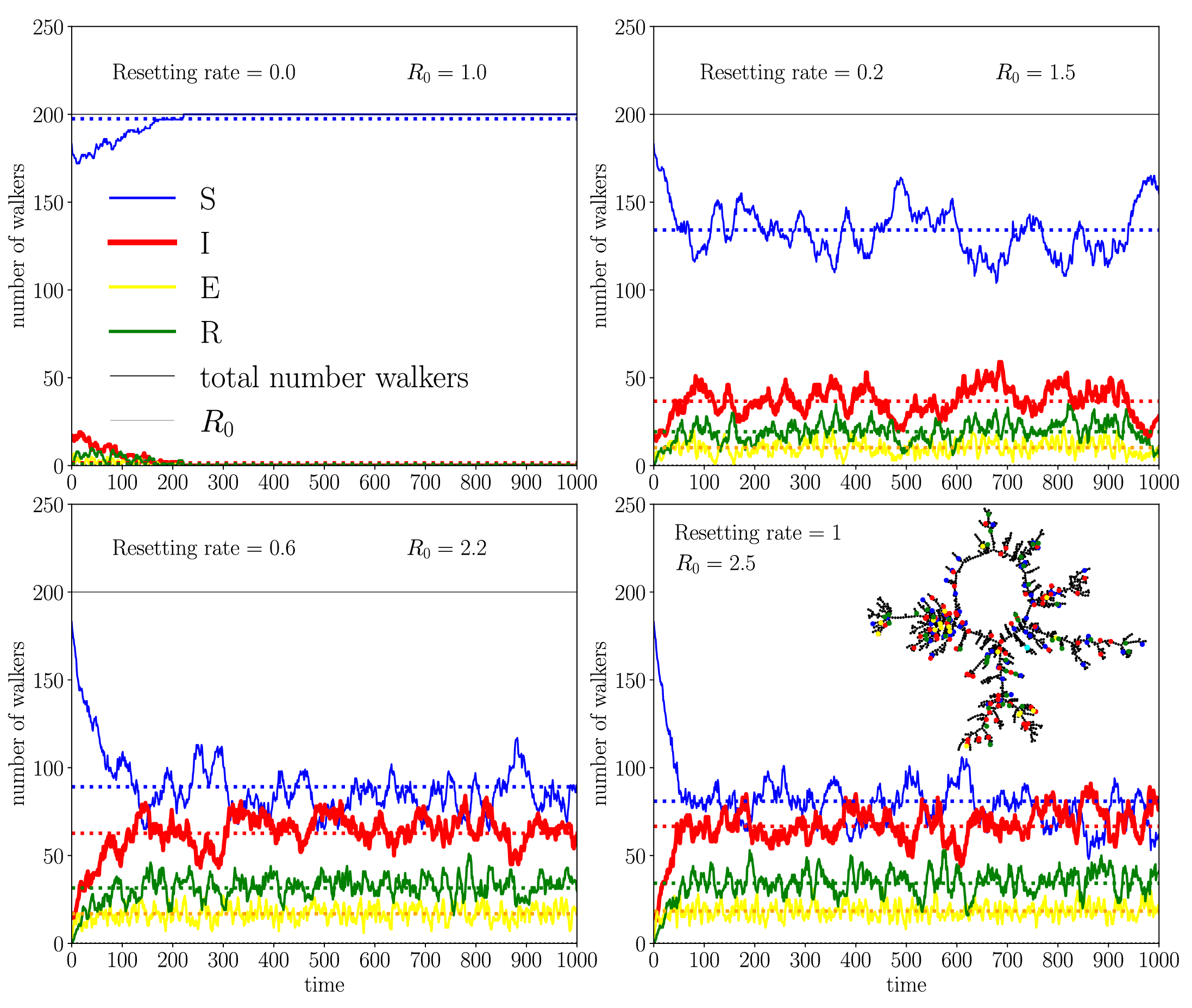} 
}
\vspace{-1mm}
\caption{Effect of resetting on the spreading for zero mortality with emergence of endemic states in a large world
Watts-Strogatz (WS) network (generated by the PYTHON NetworkX library) of $1500$ nodes with $200$ walkers. 
Colors indicate the compartments of walkers. Compartmental sojourn times are Gamma distributed with 
$\langle t_I \rangle : \langle t_R \rangle : \langle  t_E \rangle = 4 : 2: 1$, which can be identified in the plots, corroborating (\ref{endem}) for all considered resetting rates $p$. The infection state of the graph at runtime $1000$ is exhibited by the inset. The basic reproduction number ${\cal R}_0$ is monotonously increasing with $p$.}
\label{fig1}
\end{figure*}
\begin{figure*}[t!]
\centerline{
\includegraphics[width=0.92\textwidth]{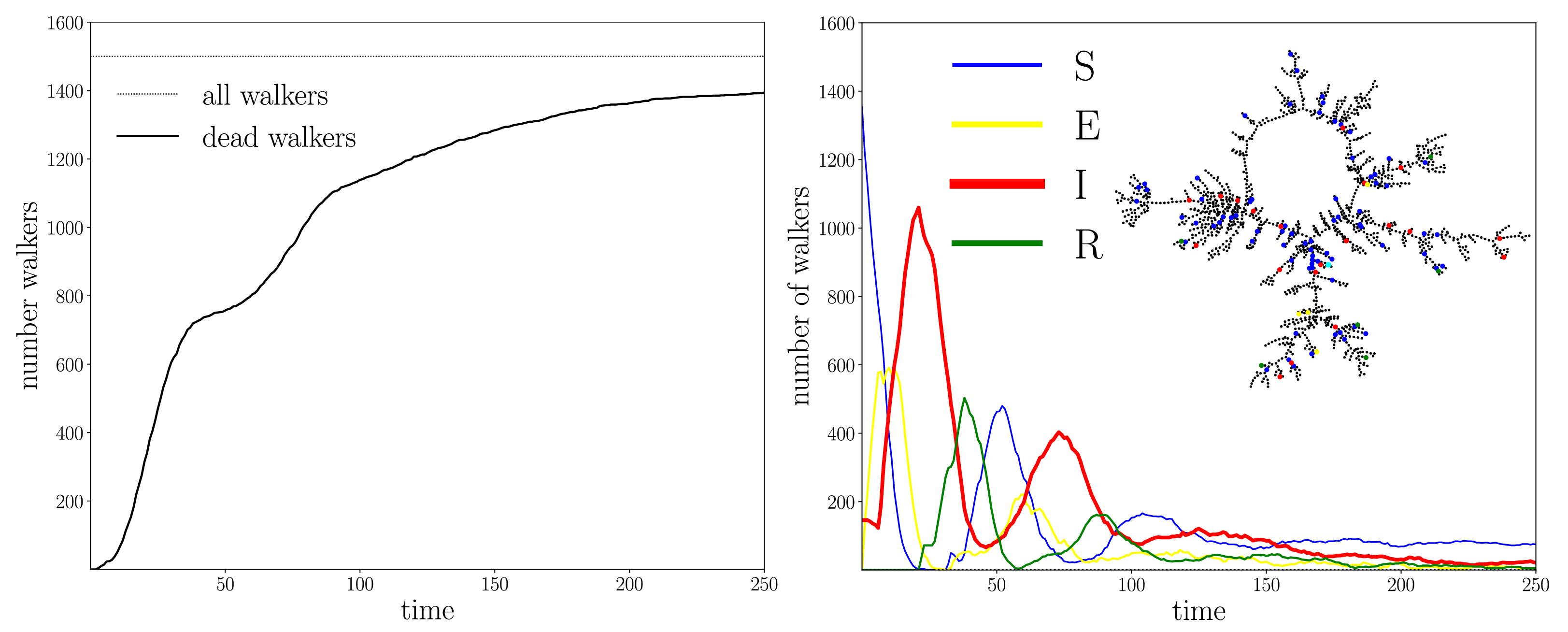} 
}
\caption{Spreading with high mortality and resetting in the WS graph of Figure \ref{fig1} for resetting probability $p=0.6$. The inset shows the infection state of the graph at runtime $t=250$  (D walkers are invisible) with eventually only about 100 survived walkers out of $1500$. We use the same color code as in Figure \ref{fig1}. The right frame depicts the epidemic wave and left frame the evolution of the cases of death.}
\label{fig2}
\end{figure*}
\begin{equation}
 \label{SCEIRS-D-model_zero-mortality}
 \begin{array}{lll}
 \displaystyle \frac{d S(t)}{dt} = & - {\cal A}(t) + \big\langle {\cal A}(t-t_E-t_I-t_R) \big\rangle & \\[1ex]
 &+  J_0 \big\langle \delta(t-t_I-t_R)\big\rangle +  R_0\big\langle \delta(t-t_R) \big\rangle &  \\[1ex]
 \displaystyle \frac{d E(t)}{dt} = & {\cal A}(t) - \big\langle {\cal A}(t-t_E) \big\rangle &\\[1ex] 
\displaystyle \frac{d J(t)}{dt}  = &\big\langle {\cal A}(t-t_E) \big\rangle  - \big\langle {\cal A}(t-t_E-t_I)\big\rangle - J_0 \big\langle \delta(t-t_I) \big\rangle   & \\[1ex]
 \displaystyle \frac{d R(t)}{dt} = & \big\langle {\cal A}(t-t_E-t_I) \big\rangle  +J_0 \big\langle \delta(t-t_I) \big\rangle & \\[1ex]
 & \displaystyle  -J_0 \big\langle \delta(t-t_I-t_R) \big\rangle    &\\[1ex]
 & \displaystyle  - R_0\big\langle \delta(t-t_R) \big\rangle - \big\langle{\cal A}(t-t_E-t_I-t_R) \big\rangle  & 
 \end{array}
\end{equation}
with $S(t)+E(t)+J(t)+R(t) =1$. 
In order to derive the endemic equilibrium, it is convenient to work with Laplace 
transformed (\ref{SCEIRS-D-model_zero-mortality}), where
${\hat f}(\lambda) = \int_0^{\infty} f(t) e^{-\lambda t} {\rm d}t $  is the LT of $f(t)$. We use the limit value theorem
$ f(\infty) = \lim_{\lambda \to 0} \lambda {\hat f}(\lambda)$ to obtain the constant asymptotic values of the endemic equilibrium as [22]
\begin{equation}
\label{endem}
\begin{array}{clr} 
\displaystyle  S_e & = \displaystyle  \frac{1}{{\cal R}_0} , &   {\cal R}_0=
\beta \langle t_I \rangle ,\\ \\
\displaystyle   E_e & = \displaystyle  \frac{{\cal R}_0-1}{{\cal R}_0} \frac{\langle t_E \rangle }{\langle T \rangle}  & \\ \\
\displaystyle  J_e & = \displaystyle  \frac{{\cal R}_0-1}{{\cal R}_0} \frac{\langle t_I \rangle}{\langle T \rangle}  &\\ \\
\displaystyle  R_e & =  \displaystyle  \frac{{\cal R}_0-1}{{\cal R}_0} \frac{\langle t_R \rangle}{\langle T \rangle} .& 
\end{array}
\end{equation}
The endemic equilibrium is independent of the initial conditions,
where $\displaystyle  \langle T \rangle = \langle t_E+t_I+t_R \rangle$ and ${\cal A}_e = \frac{{\cal R}_0-1}{{\cal R}_0}\frac{1}{\langle T \rangle}$. (\ref{endem}) exists for 
${\cal R}_0=\beta \langle t_I\rangle >1$, which also is the spreading condition of the disease. 
${\cal R}_0$ indeed is the basic reproduction number. In (\ref{endem}) $\big\langle t_{E,I,R}\big\rangle = 
\int_0^{\infty} \tau K_{E,I,R}(\tau){\rm d}\tau$ stand for the mean compartmental sojourn times, assuming here their finiteness. 
Relations (\ref{endem}) generalize the classical result [1] to arbitrary waiting time distributions and multiple compartments.
Here we consider Gamma distributed waiting times due to the high flexibility of Gamma distributions to 
adopt the behaviors of a wide range of real world diseases (see e.g., [22], [23] for details).

\section{Random walk simulations with resetting}
\label{random_walk}
We assume that each walker navigates for discrete times independently on an ergodic network [25], [26].
In order to describe the random walk of each walker, 
we denote with $i=1,\ldots N$ the nodes of the network and introduce the symmetric $N\times N$ adjacency matrix $(A_{ij})$, where $A_{ij}=1$ if the pair of nodes $i,j$ is connected by an edge, and $A_{ij}=0$ if the pair is disconnected. 
Further, we assume $A_{ii}=0$ to avoid self-connections of nodes. We restrict our analysis to undirected networks, where edges have no predefined direction and the adjacency matrix is symmetric. The degree $k_i$ of a node $i$ counts the number
of its neighbor nodes (connected with $i$ by edges).
Each walker performs independent Markovian steps between connected nodes. 
 The steps from a node $i$ to one of its $k_i=\sum_{j=1}^NA_{ij}$ neighbor nodes are chosen with probability $1/k_i$, 
 leading for all $Z$ walkers to the same transition matrix, namely [26]-[28]
\begin{equation}
\label{markovian walk}
\Pi(i \to j) = \frac{A_{ij}}{k_i} , \hspace{0.5cm} z=1,\ldots, Z , \hspace{0.5cm} i,j=1,\ldots, N ,
\end{equation}
which is by construction row-normalized $\sum_{j=1}^N\Pi(i \to j) =1$. In addition, we relocate (`reset') the walkers at each time instant to
randomly chosen nodes with a certain probability $p$. This modifies the transition matrix of the steps for each walker to 
\begin{equation}
\label{RW_resetting}
W_{i \to j} =  q\Pi(i \to j) + p R_j , \hspace{1cm} p+q = 1 ,
\end{equation}
where in our simulations we have uniform resetting probabilities $R_j=\frac{1}{N}$ to each node of the network. (\ref{RW_resetting}) introduces long-range journeys into the random walks,
and the spreading behavior is modified compared to local walks (\ref{markovian walk}). Stochastic resetting (SR) is a fundamental process in nature where dynamical systems are reset to the initial or randomly chosen states. SR occurred only a decade ago in the literature [29]
and has meanwhile launched a myriad of models and opened a wide interdisciplinary field, e.g., [30]-[33] (and many others).

\section{Results and discussion}
\label{discussion_results}
In Figure \ref{fig1}, we depict the simulated time evolution of compartmental populations (absolute numbers of walkers) under the influence of resetting for some values of relocation probability $p$ and zero mortality. The independent motion of each walker is governed by (\ref{RW_resetting}).
The parameters are such that no spreading occurs without resetting with ${\cal R}_0=1$ where the disease is eventually extinct (left upper frame). Increasing $p$ introduces more long-range displacements where the number of contacts of S and I walkers and hence infection rates with basic reproduction numbers ${\cal R}_0$ increase. The disease is spreading from $p = 0.2$ with monotonously increasing endemic values $E_e,J_e,R_e$ and ${\cal R}_0$ with $p$.
Our simulations corroborate (\ref{endem}), i.e., the ratios of the observed endemic values correspond to the ratios of mean compartmental sojourn times. 
We determined ${\cal R}_0$ in the simulations from the first equation of (\ref{endem}).

We assumed in our mean field model, a simple mass-action law for the infection rates (\ref{infection_law}), leading with (\ref{SCEIRS-D-model_zero-mortality}) to
the endemic states (\ref{endem}). These endemic values are in excellent agreement with the large-time asymptotics obtained from the random walk simulations (see Figure \ref{fig1}). This remains true when the random walks of the individuals are subjected to resetting, which in the large time limit affects only the macroscopic transmission coefficient $\beta$. 
These observations suggest that random walks indeed offer suitable microscopic pictures of the corresponding spreading dynamics.

Animated simulation-videos on Watts-Strogatz graphs 
can be launched online by clicking on the slanted text for a case
\href{https://drive.google.com/file/d/1CAsCVIX4Jk97UPg9lQVRCrR_igoX0qr9/view}{\it without mortality and no resetting} (see relation (\ref{SCEIRS-D-model_zero-mortality})). A further animation video of the spreading under resetting ($p=0.6$) with graph of Figure \ref{fig1} and similar parameters
\href{https://drive.google.com/file/d/1cslrP7kj4iJGiI-6zoDs8H4kMa6snYVX/view}{\it includes mortality} 
(see relations (\ref{SCEIRS-D-model}), (\ref{SCEIRS-D-model-mortality})).
Simulation (Python) codes with parameters and further details can be obtained upon request or consult our website \href{https://sites.google.com/view/scirs-model-supplementaries/accueil}{\it supplementary materials}.

\section{Conclusion and future work}
\label{Conclusions}

We proposed a multiple random walker's epidemic compartment model, which accounts for mortality: An infected walker may die during the period of its infection. We excluded demographic birth and death processes. 
The compartmental sojourn times were considered to be independent random variables drawn 
from specific (here Gamma-) distributions. By including stochastic resetting into the random walks, 
in which walkers are relocated to random positions, 
we are able to mimic the effects of long-range voyages on the spread of the disease.
By considering zero mortality, we observed that the macroscopic compartment model (endemic states (\ref{endem})) remains true for any resetting rate $p$, where the macroscopic transmission coefficient $\beta$ is monotonously 
increasing with the resetting rate. Increasing numbers of long-range journeys may drive the basic reproduction number to values above one, 
which launches the spreading of the disease. It follows that measures reducing long-range voyages can be an effective way to block
the propagation of an epidemic. 
The results of the simulations suggest that in all cases, above equations
(\ref{endem}) for the endemic states remain valid and capture well the large time asymptotics.
We conclude that our random walk approach of multiple walker's navigating independently in a complex network is a powerful tool to capture the dynamics of the epidemic spreading.
Our model can be generalized in several directions, for instance, to vector-borne transmission 
pathways [23] or assuming non-monotonous infection rates (different
from simple mass-action-laws) for which under certain conditions the endemic
equilibrium exhibits bifurcations, allowing for emergence of chaotic attractors
[34]. 
A promising direction is to account for infection rates beyond the present mass-action law (\ref{infection_law}) by including information of the network topology and the random walk. Introduction of individual navigation rules for specific walkers can be of interest as well.


\begin{thebibliography}{100}

\bibitem{KermackMcKendrick1927} W. O. Kermack and A. G. McKendrick, A contribution to the mathematical 
theory of epidemics, Proc. Roy. Soc. A 115, 700–721, 1927.

\bibitem{Soper1929} H. E. Soper, The interpretation of periodicity in disease prevalence, J. Royal Statistical Society 92, 34-61, 1929.

\bibitem{LiuHeathcote1987} W. M. Liu, H. W. Hethcote and S. A. Levin, Dynamical behavior of epidemiological
models with non-linear incidence rate. J. Math. Biol. 25, 359–380, 1987.

\bibitem{Li-etal1999} M. Y. Li, J. R. Graef, L. Wang and J. Karsai, Global dynamics of a SEIR model with
varying total population size. Math. Biosci. 160:191–213, 1999.

\bibitem{Anderson1992} R. M. Anderson and R. M. May, Infectious Diseases in Humans, (Oxford University
Press, Oxford), 1992.

\bibitem{Martcheva2015} M. Martcheva, An Introduction to Mathematical Epidemiology, Springer, 2015.

\bibitem{Harris2023} J. E. Harris, Population-Based Model of the Fraction of Incidental
COVID-19 Hospitalizations during the Omicron BA.1 Wave in the United States, COVID 3(5), 728-743, 2023.
Doi: 10.3390/covid3050054

\bibitem{Satoras-Vespignani-etal2015} R. Pastor-Satorras, C. Castellano and P. Van Mieghem, A. Vespignani,
Epidemic processes in complex networks, Rev. Mod. Phys. 87, 925-979, 2015.

\bibitem{Pastor-SatorrasVespignani2001} R. Pastor-Satorras and A. Vespignani, Epidemic dynamics and endemic states in complex networks, Phys. Rev. E 63, 066117, 2001.

\bibitem{OkabeShudo2021} Y. Okabe Y and A. Shudo, Microscopic Numerical Simulations of Epidemic Models on Networks. Mathematics 9, 932, 2021.

\bibitem{Barabasi2016} A.-L. Barab\'asi, Network Science. Cambridge University Press, Cambridge, 2016.

\bibitem{BarabasiAlbert1999} A.-L. Barab\'asi and R. Albert, Emergence of Scaling in Random Networks, Science 286, 509, 1999.

\bibitem{Barrat-etal2008} A. Barrat, M. Barth\'elemy and A. Vespignani, Epidemic spreading in population networks, In: Dynamic Processes on Complex Networks, pp. 180 -- 215, Cambridge University Press, 2008, DOI: 10.1017/CBO9780511791383.010 

\bibitem{Ross1996} S.M. Ross, Stochastic Processes (John Wiley \& Sons,
New York), 1996.

\bibitem{fractional_book_MiRia2019} T. M. Michelitsch, A. P. Riascos, B. A. Collet, A. F. Nowakowski and F. C. G. A. Nicolleau, Fractional Dynamics on Networks and Lattices, ISTE/Wiley, London, 2019.

\bibitem{RiascosSanders2021} A. P. Riascos and D. P. Sanders, 
Mean encounter times for multiple random walkers on networks,
Phys. Rev. E 103, 042312, 2021.

\bibitem{BesMi-etal2021}
M. Bestehorn, A. P. Riascos, T. M. Michelitsch and B. A. Collet, A Markovian random walk model of epidemic spreading,
Continuum Mech. Thermodyn. 33:1207–1221, 2021. Doi: doi.org/10.1007/s00161-021-00970-z

\bibitem{vanKampen1981} N. G. van Kampen, Stochastic processes in chemistry
and physics (North Holland, Amsterdam), 1981.

\bibitem{VanMiegem2014}
P. Van Mieghem, Exact Markovian SIR and SIS epidemics on networks and an upper
bound for the epidemic threshold, 2014, arXiv:1402.1731 

\bibitem{BasnakovSandev-etal2020} L Basnarkov, I. Tomovski, T. Sandev and L. Kocarev, Non-Markovian SIR epidemic spreading model of COVID-19, Chaos, Solitons and Fractals 160, 112286, 2022.

\bibitem{BesMiRias2022}
M. Bestehorn, T. M. Michelitsch, B. A. Collet, A. P. Riascos,
and A. F. Nowakowski, Simple model of epidemic dynamics with memory effects, Phys. Rev. E 105, 024205, 2022.

\bibitem{Granger-et-al2023} T. Granger, T. M. Michelitsch, M. Bestehorn, A. P. Riascos and B. A. Collet,
Four-compartment epidemic model with retarded transition rates, Phys. Rev. E 107 044207, 2023.

\bibitem{entropy_paper} T. Granger, T. M. Michelitsch, M. Bestehorn, A. P. Riascos and B. A. Collet, Stochastic Compartment Model with Mortality and Its Application to Epidemic Spreading in Complex Networks,  Entropy 26(5), 362, 2024.

\bibitem{BesMi2023} M. Bestehorn and T. M. Michelitsch, Oscillating Behavior of a Compartmental 
Model with Retarded Noisy Dynamic Infection Rate, 
Int. J. Bifurcation Chaos 33 2350056, 2023. 

\bibitem{RiascosSanders2021} A. P. Riascos and D. P. Sanders, Mean encounter times for multiple random walkers on networks, Phys. Rev. E 103, 042312, 2021.

\bibitem{Newman2010} M.E.J. Newman, Networks: An Introduction, Oxford University Press, Oxford, 2010.

\bibitem{NohRieger2004} J.-D. Noh and H. Rieger, Random walks on complex networks, 
Phys. Rev. Lett. 92, No. 11, 2004. 

\bibitem{fractional_book_MiRia2019}  T. M. Michelitsch, A. P. Riascos, B. A. Collet, A. F. Nowakowski and F. C. G. A. Nicolleau, Fractional Dynamics on Networks and Lattices, ISTE/Wiley, London, 2019.

\bibitem{resetting_first}
M. R. Evans and S. N. Majumdar, Diffusion with stochastic resetting, Phys. Rev. Lett. 106, 160601, 2011.

\bibitem{mi_resetting_chaos2025} T.M. Michelitsch, G. D’Onofrio, F. Polito and A. P. Riascos, 
Random walks with stochastic resetting in complex networks: A discrete-time approach,  
Chaos 35, 013119, 2025.

\bibitem{ria_resetting_chaos2025} A. G. Guerrero-Estrada, A.P. Riascos and D. Boyer, Random walks with long-range memory on networks, Chaos 35, 013117, 2025.

\bibitem{Pal_Sandev2015}
A. Pal, V. Stojkoski and T. Sandev, Random resetting in search problems. In: Target Search Problems, edited by D. Grebekov, R. Metzler, and G. Oshanin ,Springer Nature, 2024, ISBN: 978-3-031-67801-1 (arXiv:2310.12057).

\bibitem{Sandev_Ionin2025}  T. Sandev, A. Iomin, J. Kurths and L. Kocarev, Shear-driven anomalous diffusion: Memory effects and stochastic resetting, Physics of Fluids 37, 067101, 2025.

\bibitem{bes_mi_chaos2025} M. Bestehorn and T.M. Michelitsch, 
Periodic solutions and chaotic attractors of a modified epidemiological SEIS model, 
Chaos 35, 023104, 2025.

\end{thebibliography}
\end{document}